\documentclass[12pt,preprint]{aastex}
\usepackage{natbib}
\slugcomment {Accepted by ApJ}
\shorttitle{Temperature Structure of Pre-Stellar Cores}
\shortauthors{Bergin et al.}

\def\lesssim{\mathrel{\hbox{\rlap{\hbox{\lower4pt\hbox{$\sim$}}}\hbox{$<$}}}}
\def\gtrsim{\mathrel{\hbox{\rlap{\hbox{\lower4pt\hbox{$\sim$}}}\hbox{$>$}}}}
\newcommand\go{\rm $G_0$}

\newcommand\be {\begin{equation}}
\newcommand\en{\end{equation}}

\newcommand\av{\rm $A_V$}
\newcommand{\cc}{\mbox{cm$^{-3}$}}

\newcommand{\nhtwo}{\mbox{n$_{H_{2}}$}}

\def\m17{M~17}               
\def\thCO{$^{13}$CO}           
\def\CeiO{C$^{18}$O}           
\def\CtfS{C$^{34}$S}           
\def\NtwoHp{N$_2$H$^+$}        
\def\NHthree{NH$_{3}$}         

\def\m{\ts {\rm m}}

\def\swash2o{$1_{10} - 1_{01}$}             

\let\ts=\thinspace

\slugcomment{Accepted by {\it The Astrophysical Journal }}

\begin{document}

\title
{
The Thermal Structure of Gas in Pre-Stellar Cores: A Case Study of Barnard 68 
}

\author{Edwin A. Bergin\altaffilmark{1},
S\'ebastien Maret\altaffilmark{1},
Floris F.S. van der Tak\altaffilmark{2},
Jo\~ao Alves\altaffilmark{3},
Sean M. Carmody\altaffilmark{4},
Charles J. Lada\altaffilmark{5}
}

\altaffiltext{1}{University of Michigan, 825 Dennison Building, 500
Church St,
Ann Arbor, MI 48109-1042; email: ebergin@umich.edu}
\altaffiltext{2}{
Max-Planck-Institut f\"ur Radioastronomie,
Auf dem H\"ugel 69, 53121 Bonn, Germany
}
\altaffiltext{3}{
European Southern Observatory, Karl-Schwarzschild-Strasse 2, 85748
Garching, Germany
}
\altaffiltext{4}{
4824 S. 1st Street
Kalamazoo, MI 49009
}
\altaffiltext{5}{
Harvard-Smithsonian Center for Astrophysics, 60 Garden Street,
Cambridge, MA 02138
}

\begin{abstract}
We present a direct comparison of a chemical/physical model to multitransitional
observations of \CeiO\ and \thCO\ towards the Barnard 68 pre-stellar core.
These observations provide a sensitive test for models of 
low UV field photodissociation regions and offer the best constraint on
the gas temperature of a pre-stellar core.  
We find that the gas temperature of this object is 
surprisingly low ($\sim 7-8$ K), and
significantly below the dust temperature, in the outer layers (\av\ $<$ 5 mag)
that are traced by \CeiO\ and \thCO\  emission.  As shown previously,
the inner layers 
(\av\ $> 5$ mag) exhibit significant freeze-out of CO onto grain surfaces.  Because the
dust and gas are not fully coupled, depletion of key coolants in the densest layers
raises the core (gas) temperature, but only by $\sim 1$ K.  
The gas temperature in layers not traced by
\CeiO\ and \thCO\ emission can be probed by \NHthree\ emission, with 
a previously estimated temperature of $\sim 10-11$ K.   To reach these temperatures in the
inner core requires an order of magnitude reduction in the gas to dust coupling rate.
This potentially argues for a lack of small grains in the densest gas, presumably due to grain
coagulation.  
\end{abstract}


\section{Introduction}
\label{sec_intro}

Several phases of the star formation process have been 
observationally identified and characterized, 
beginning with a centrally concentrated core of molecular gas 
that collapses to form a star surrounded by a disk.  
Indeed, it has been the isolation of objects that have not yet formed 
stars -- pre-stellar cores -- which has allowed us to probe  
the earliest stages of star formation 
\citep[see ][]{andre_ppiv, alves_b68, lee_myers_survey}.
Of particular importance are the density and temperature structure as
these parameters are fundamental to estimating the mechanisms
via which cores are supported against gravitational collapse.  Moreover
knowledge of the physical parameters is a pre-requisite for 
subsequent chemical and kinematical studies \citep[e.g.
][]{vdt_l1544ion}.

Estimates of the density profile of pre-stellar cores have become
quite commonplace through observations of dust emission
or absorption.  Initial studies have suggested that 
pre-stellar cores have flat density profiles in the center, falling
off as a steep power law in outer layers \citep[e.g.][and references therein]{andre_ppiv}.
Less information exists regarding the temperature structure.
This is despite the importance of thermal pressure for the stability
of pre-stellar cores with little or no turbulent  support 
\citep{dickman_bok, lada_b68}.
Initial studies assumed isothermal structure with 
equivalent gas and dust temperatures 
\citep[as a result of the anticipated thermal coupling at 
\nhtwo\ $> 10^5$ cm$^{-3}$;][]{burke_hollenbach}.
More recently, several groups have investigated the thermal structure of  
dust grains heated by the normal interstellar radiation field 
\citep{evans_td, zucconi_td, bianchi_td}.
 These studies predict a dust temperature gradient that peaks at 15 -- 17 K 
towards the core edge and declines to near 7 - 8 K at the center in cores with central
densities of $\sim 10^6$ \cc .
In this case the actual gas density profile, when estimated via dust continuum emission,
could be much steeper than derived assuming constant temperature, for example as steep 
as $r^{-2}$ \citep{evans_td}.

Despite the effort placed towards the determination of the dust 
temperature profile little has been done to characterize the temperature
profile of the gas, which is the dominant component.   
For cores bathed within the average interstellar radiation field (ISRF), 
theory predicts that the gas temperature will rise towards the
cloud edge due to the direct heating of the gas via the photo-electric effect;
in shielded gas heating by cosmic rays and dust-gas coupling is expected to 
dominate \citep{goldsmith_tg, galli_tg}.

Excellent tracers of the gas temperature are the 22 GHz inversion
transitions of NH$_3$ and cm/mm-wave transitions of H$_2$CO.
Emission from H$_2$CO has been used to
probe the gas temperature in pre-stellar cores by \citet{young_tgas}
with a finding that this species is heavily depleted in abundance in the
center, limiting its value as a probe to outer layers.
NH$_3$ does not appear to exhibit significant gas-phase freeze-out
\citep{tafalla_dep}
and should directly trace T$_{gas}$ deep inside cores.
Indeed the gas kinetic temperature derived using ammonia 
is $\sim 9 - 11$ K, consistent with cosmic ray heating, but 
above that derived for the dust in similar objects    
\citep{tafalla_dep, hotzel_nh3, lai_nh3, evans_td, zucconi_td}.   This is predicted by
models of the thermal balance for cores where the central density is $\sim 10^5$ \cc
; in cores with higher densities (e.g. L1544) the gas temperature is expected to drop
\citep{galli_tg}.
However, NH$_3$ is limited as a probe of the full thermal structure
because its emission does not probe the outer layers of the core
seen in other tracers \citep[e.g. C$^{18}$O, CS, H$_2$CO; ][]{zhou_cs,
tafalla_dep, young_tgas}.  
Thus, while detailed estimates of the
dust temperature profile in pre-stellar cores exist and 
have been compared to observational data, less 
observational constraints exist for the gas temperature profile.

This paper presents a detailed study of the gas temperature structure in the Barnard 68 
pre-stellar core.  This well-studied object presents a unique laboratory for study as its
density and extinction structure is well determined via near-infrared extinction mapping
techniques \citep[see][and references therein]{alves_b68}. This allows for good
constraints to be placed on the line of sight structure in chemical abundance.
Observations have shown that this, and other, pre-stellar
cores are dominated by selective freeze-out
\citep{tafalla_dep, bergin_b68}, wherein some species (e.g. CS and CO) only trace
outer layers and others (\NtwoHp , \NHthree ) trace deeper into the cloud.  With
knowledge of the chemical structure obtained by detailed models we can determine which
tracer is probing a given layer and use multi-molecular studies to reconstruct the physical
structure of the cloud.  Thus CO and its isotopologues
can be used to constrain the gas
temperature in the outer low density layers, while \NHthree\ 
probes denser gas.  This enables us to perform a more detailed
examination of the gas temperature structure in the earliest stages of star
formation than has been done previously.

In \S 2 we present our observations and in \S 3 we discuss the model used to compare
to the data.  Section 4 compares theory to observations and 
in \S 5--6 we discuss the implication of and summarize our results.

\section{Observations and Results}
\label{sec:obs}

The J=1-0  (109.78218 GHz) and J=2-1 (219.560319 GHz) transitions of \CeiO\
were observed towards B68 
($\alpha$ = 17$^h$22$^m$38\fs 2 and $\delta = -$23$\arcdeg$49$'$34\farcs 0; J2000)
during April 2000 and 2001 using the
IRAM 30m telescope. The entire core was mapped using  12$''$ sampling  
with the  half-power beam width of 22$''$ at 110 GHz and 11$''$ at 220 GHz.
System temperatures for the 3mm observations were typically $\sim$160--190 K with
$\sim$350--450 K at 1mm and integration times were typically 2--6 min.
Pointing was checked frequently with 
an uncertainty of $\sim 2''$.  The data were taken using frequency switching and
calibrated using the standard chopper wheel method and are presented here
on the T$_{mb}$ scale using standard calibrations from IRAM documentation.  
For the analysis the \CeiO\ J=2-1 observations were convolved to the lower resolution
of the J=1-0 transition.
 The J=2-1 data were reduced by fitting the spectra with Gaussians and in cases where no
emission is observed fixing the width and position of the line using a centroid determined
from J=1-0 observations.
The J=1-0 observations have been published previously by
\citet{bergin_b68} and \citet{lada_b68}.

Observations of \thCO\ J=2-1 (220.3987 GHz; $\theta_{MB} \sim 33''$), \thCO\ J=3-2 (330.588 GHz
$\theta_{MB} \sim 22''$), and \CeiO\ J=3-2 (329.330 GHz; $\theta_{MB} \sim 32''$) 
were obtained at the 10.4m Caltech submillimeter Observatory (CSO) during
September 7-9, 2003.  Data were taken using position switching with an offset
of 6$'$; larger reference position offsets for the position switched observations were 
examined with no difference found in the results.
Pointing was checked frequently with a typical uncertainty of 3--4$''$.
Typical system temperatures range from $\sim$400 -- 500 K for the J=2-1 transition
($t_{int} \sim 12-20$ s) and $\sim
600-700$ K ($t_{int} \sim 40$ s) 
for the J=3-2 transitions.   The \thCO\ transitions were observed over
numerous positions in the core with the goal of obtaining data 
corresponding to a range of total column (as estimated from the extinction map).
Observations of \CeiO\ J=3-2 were taken towards two positions in the core with relative
offsets of $\Delta\alpha = 0$, $\Delta\delta = 0$ and $\Delta\alpha = 24''$,
$\Delta\delta = -48''$.  No line is detected at either position with a 1$\sigma$ rms of
$\Delta T_A^*$ = 0.23 K and 0.25 K, respectively in 0.05 MHz channels.
An additional CSO search for  \CeiO\ J=3-2 emission was performed on
May 27, 2005 (by D. Lis) with no emission detected to a level of 
0.3 K (1$\sigma$) in 0.05 MHz channels.

All data were calibrated to the T$_A^*$ scale using the standard chopper wheel method.  
To compare the CSO data to IRAM 30m observations we placed the data on the
T$_{MB}$ scale.  For \CeiO\ and \thCO\ J=3-2 observations of Mars on Sep. 7, 2003
($\theta_{Mars} = 24.6''$) were compared to a
thermal model (M. Gurwell, priv. comm.) to estimate a main beam efficiency of 55\%.
For J=2-1 we adopt a main beam efficiency of 60\% (D. Lis, priv. comm.).  
To test this cross-telescope
calibration we also observed \thCO\ J=2-1 in Lynds 1544 with the CSO  and compared this
to similar data obtained with the IRAM 30m (kindly provided by M. Tafalla).
We find that the IRAM 30m data (when convolved to the same resolution as CSO) agrees to
within 20\%. The IRAM data shows stronger emission, which could be
attributed to a 10--20\% contribution from the error beam of the larger IRAM telescope
\citep[see][]{bensch_iram}.  Given these differences, we have assigned a calibration
uncertainty of 20\% to all 230 and 345 GHz data.   \CeiO\ J=1-0 data
are assigned a
calibration uncertainty of 10\%.

Fig.~\ref{fig:time_data} presents the observational data as a function
of visual extinction.  In this plot (and in subsequent plots) the visual
extinction was computed for a 24$''$ beam on the same grid as the
molecular observations using the infrared extinction data of
\citet{alves_b68}.  Thus molecular emission is paired with the
visual extinction for each position on a point by point basis.  
 Also shown in Fig.~\ref{fig:time_data}a is a weighted 1 A$_V$ binned average of the
C$^{18}$O J=2-1/J=1-0 line ratio.  This observed ratio shows increased scatter for
\av\ $< 5$  mag, but the average line lies close to a ratio of $\sim 0.9$ even at low
extinction.
Thus, one 
simple result can be extracted from this data.  Because the \CeiO\
emission is optically thin the (2--1)/(1--0) emission line ratios can be
used to estimate the excitation temperature (which due to the low dipole
moment of \CeiO\ should be relatively close to the gas kinetic
temperature).  The (2--1)/(1--0) emission line ratio of $\sim$0.9 
corresponds to an excitation temperature of 7.6 K, in agreement
with previous estimates \citep{hotzel_b68}.

From the \thCO (2-1)/\CeiO
(2-1) line ratio of 0.53 we also estimate the \CeiO\ J=2--1 emission
is nearly optically thick with $\tau \sim 0.8$.
Thus, the gas in the
region probed by \CeiO\ is colder than the 
canonical 10 K temperature of typically assumed for low mass cores.

\section{Model}
\label{sec:model}

\subsection{Physical Model}

 The density structure of Barnard 68 has been tightly constrained via near-infrared extinction
mapping techniques by \citet{alves_b68} and is consistent with that of a pressure confined
self-gravitating isothermal sphere \citep[a Bonnor-Ebert sphere;][]{bonnor, ebert}. 
The extinction measurements show a decline from 30 mag to 1 mag within a radial
distance of $\sim 120''$ corresponding to $\sim 0.067$ pc at a our adopted distance of
125 pc (see Fig.~2 in that paper).  Thus  with a central density of 
n$_{\rm HI + 2H_2} \sim 6 \times 10^5$ cm$^{-3}$\footnote{
This central density is derived from the Bonnor-Ebert fit with a value of
$\zeta_{max} = 6.9$.  The central density, $n_c$ is proportional to $n_c \propto T\zeta_{max}^2/D^2$, 
where $D$ is the distance and $T$ is the temperature.  We have adopted values of $n_c$ from
\citet{alves_b68}, who assume the gas and dust temperature is 16 K.
Based on measurements from \NHthree , this temperature is believed to be too high \citep{hotzel_nh3,
lai_nh3}.  
Since the distance is also uncertain we use $n_c = 6 \times 10^5$ \cc .  We have examined models with a density  change of $\sim 40$\%
 finding no difference in our results.}, 
and assuming spherical structure, the
density at the core edge is quite high $> 2 \times 10^{4}$ cm$^{-3}$.  With the sharp
fall off characteristic of pressure confined Bonnor-Ebert spheres there is little mass 
at lower densities.  
In Fig.~\ref{fig:fuv_model}a we show the adopted density structure as a function of visual
extinction, which are the two key parameters used as inputs to the chemical model discussed
below.

\subsection{Chemical/Thermal Model}

The chemical model adopted in this work is the gas-grain network described previously
in \citet{blg95} and \citet{bl97}.  We have used a smaller chemical network that
focuses exclusively on the formation of CO and simple oxygen bearing molecules.
This network was used by \citet{bmn98} and has been tested against the larger network
for consistency.  The network includes the effects of molecular
freeze-out onto grain surfaces and desorption via cosmic ray impacts
\citep{hh93, bringa} and ultraviolet photodesorption.
 The rate of UV photodesorption is given by R $= 2 \times 10^{-10} YG_0exp(-1.8A_V)$
molecules$^{-1}$ s$^{-1}$ \citep{boland_dejong, draine_salpeter}, where $Y$ is the yield
(probability of ejection from mantle per photon absorbed).  We have adopted $Y = 10^{-3}$
in our work.\footnote{
Recent laboratory measurements of CO photodesorption suggest a probability well below our
assumed value, with $Y < 1.9 \times 10^{-5}$ \citep{fleur_phdt}.  
We have examined a model with this yield and find only a small (10--15\%) 
effect on the \thCO\ and \CeiO\ emission.}

The reaction network has been 
expanded to include the effects of isotopic
fractionation using the rates and formalism described in \citet{langer_frac} and
isotopic selective photodissociation using the self-shielding rates of \citet{vdb88}.
For solving the chemical equations we use the DVODE algorithm \citep{vode}. 

The binding energy of molecules to grain surfaces will influence the
evaporation rates.  In this work we assume a CO binding energy
of 1181 K, which is equivalent to CO bound on water ice 
\citep{collings_cobind}.\footnote{Note that this value is substantially
lower than previous measurements \citep{sandford_cobind}.}   All
other binding energies are based upon the work of \citet{hh93}.

To examine the thermal balance we use a variant of the chemical/dynamical model
presented by \citet{bergin_cform}, which directly linked the time
dependent chemistry to the thermal balance.  Heating contributors
included cosmic rays, X-rays, and the photoelectric effect 
with cooling arising from
atomic fine structure and molecular rotational transitions.  References for the
adopted formalism are provided in Table~2 of that paper.  Some changes
are adopted in the present work.  
We use the formalism for cosmic ray heating from
\citet{wolfire} and have included the cooling contribution from \thCO\ using the 
tables (for CO) of \citet{nlm95}.  We have also adopted the gas-dust coupling rate from
\citet{th85}.   For the dust temperature we use the analytical expressions from
\citet{zucconi_td}, which used the \citet{mathis_isrf}, taking into account the latest
adjustments from \citet{black94}.  
We did account for variation of the dust
temperature with a lower external UV field by adding additional layers of
extinction based upon the following expression: $A_{V,add} =
log(G_0)/(-2.5)$.  This expression assumes that the heating effects of
the UV field decay as $e^{-2.5A_V}$ and is likely an overestimate (to
below a factor of $\sim$1.5) as longer
wavelength photons also contribute to dust heating.  However, it provides a
crude estimate of the effects of a variable UV field.
The dust temperature is assumed to be constant with
time, while the gas temperature will change as the chemistry evolves.

Since B68 is a cold pre-stellar object bathed in the
normal interstellar radiation field (ISRF) 
we have not coupled the thermal balance directly to the chemistry.
Rather the chemistry is calculated assuming a constant temperature of 10 K and
the thermal balance is determined using the results from the time-dependent 
chemical model.  Both the chemistry and the thermal balance assume the density
profile (derived from dust absorption) is constant with time.  
In this fashion, for each time, we derive chemical
abundances as a function of cloud depth and then compute the resulting thermal
balance.  This separation is reasonable given that 
the gas-phase chemistry is rather insensitive to the few degree changes as a function
of depth that are found from our thermal balance 
calculations. The primary temperature
dependence in the chemistry 
arises from the isotopic fractionation of CO at the cloud edges,
where the column is small.  For example, 2--3 K changes in temperature
show 20\% differences at the peak \CeiO\ abundance.

Since we have adopted a density profile that is constant with time we have
assumed that the chemistry has previously evolved at lower density
to the point that all carbon resides in CO and
oxygen is locked on grains in the form of water ice.   
In Table~1 we provide a list of assumed initial abundances.

The variables in our calculations are (1) the overall intensity of FUV radiation
field, parameterized by G$_0$, the FUV flux relative to that measured for
the local ISM by \citet{habing68}.  This influences the 
photodissociation rate (at what depth a given CO isotopologue appears) and the
amount of photoelectric heating, which dominates the heating at low \av .
(2) Several values of the primary cosmic ray ionization
rate were examined between $\zeta = 1.5 - 6.0 \times 10^{-17}$ s$^{-1}$.
CO is assumed to be pre-existing in our calculations and
changes in the cosmic ray ionization rate by factors of a few does not change
its chemistry appreciably, which is dominated by the grain freeze-out.  
It does however change the cosmic ray heating rate, 
which is important for gas heating at high extinction.
Because the cosmic ray desorption rate is 
uncertain we assumed a constant rate in our analysis  based in the formalism of \citet{hh93} and the 
binding energy given above.
(3) Time is also a variable.   As the chemistry evolves the effects of
freeze-out alters the resulting emission.
(4)  The final variable is dust-gas heating/cooling rate.
The dust-gas thermal coupling  expression from
\citet{th85} is given as:

\[
\Gamma_{gas-dust} = 3.5 \times 10^{-34}n^2
\delta_{d}T_{gas}^{0.52}k(T_{dust}-T_{gas})\;\;{\rm ergs\; cm^{-3} s^{-1}}.
\]

\noindent In this expression
$\delta_d$ is the ratio of the dust abundance in the modeled
region to that of the diffuse ISM. Effectively this is a measure of the surface
area trapped in more numerous small grains, which are more efficient 
in thermally coupling to the gas.     When $\delta_d$ is varied in our models we make
no correction for this in the chemistry, where it can alter the timescales. In general, chemical models assume that large
grains are responsible for freeze-out (which is the dominant effect in this calculation).  This assumes that cosmic rays can effectively thermally cycle small grains and leave the  surface bare
\citep{leger85}.  Thus the change in the surface area or number density of the large 1000 \AA\ grains from the loss of small grains is assumed to be small. 

Our analysis proceeds in the following fashion.
The adopted n$_{H_2}$($r$) profile and T$_{gas} =$
T$_{dust} =$ 10 K is used as inputs to the time-dependent gas-grain chemical
model.   The results from the chemical model are used as inputs to the thermal
balance calculation which provides T$_{gas}(r)$.  The abundance and gas
temperature structure are incorporated as inputs to the 
one-dimensional spherical  
Monte-Carlo radiation transfer model
\citep{ashby02}.   This model self-consistently solves for the molecular emission accounting
for effects of sub-thermal excitation, radiation trapping, and pumping by dust continuum
photons.  The latter is unimportant for B68. 
Additional inputs to this model are the
velocity line width and line of sight velocity field.  
The line width includes contributions from the thermal and turbulent widths.
In our iterations
we have assumed a static cloud with the
turbulent contribution to the linewidth included that
increases with radius from the core center \citep[as required by
observations; ][]{lada_b68}.
The model emission is
convolved to the observed angular resolution for each transition
(assuming a distance of 125 pc)
and onto a grid sampled every 12$''$.
The model emission is then placed onto $\int T_{mb}\Delta v$ vs. \av\ plots 
shown in \S 4.  To compare the velocity width
to the data we fit spectra with Gaussians.
To be successful a model must reproduce both the dependence of $\int T$dv 
with \av\ for C$^{18}$O J=1-0, \thCO\ J=2-1, J=3-2, the \CeiO\ J=2-1/1-0
integrated emission ratio, and $\Delta$v with radius for 
C$^{18}$O J=1-0 and \thCO\ J=2-1.  

\section{Analysis}

\subsection{Time Dependent Chemistry and Gas Temperature}
\label{sec:compobs}

In our analysis it became clear that we cannot 
reproduce the observed emission with the core exposed to the standard ISRF
(\go\ = 1).  Instead these data require colder outer layers 
and a reduced UV field, \go\ $\sim 0.2$.
For our initial discussion of the overall gas temperature structure in cores
dominated by freeze-out, we use the model with the reduced UV field.
In \S~\ref{sec:uvfield} we demonstrate this requirement.

\begin{figure}
\plotone{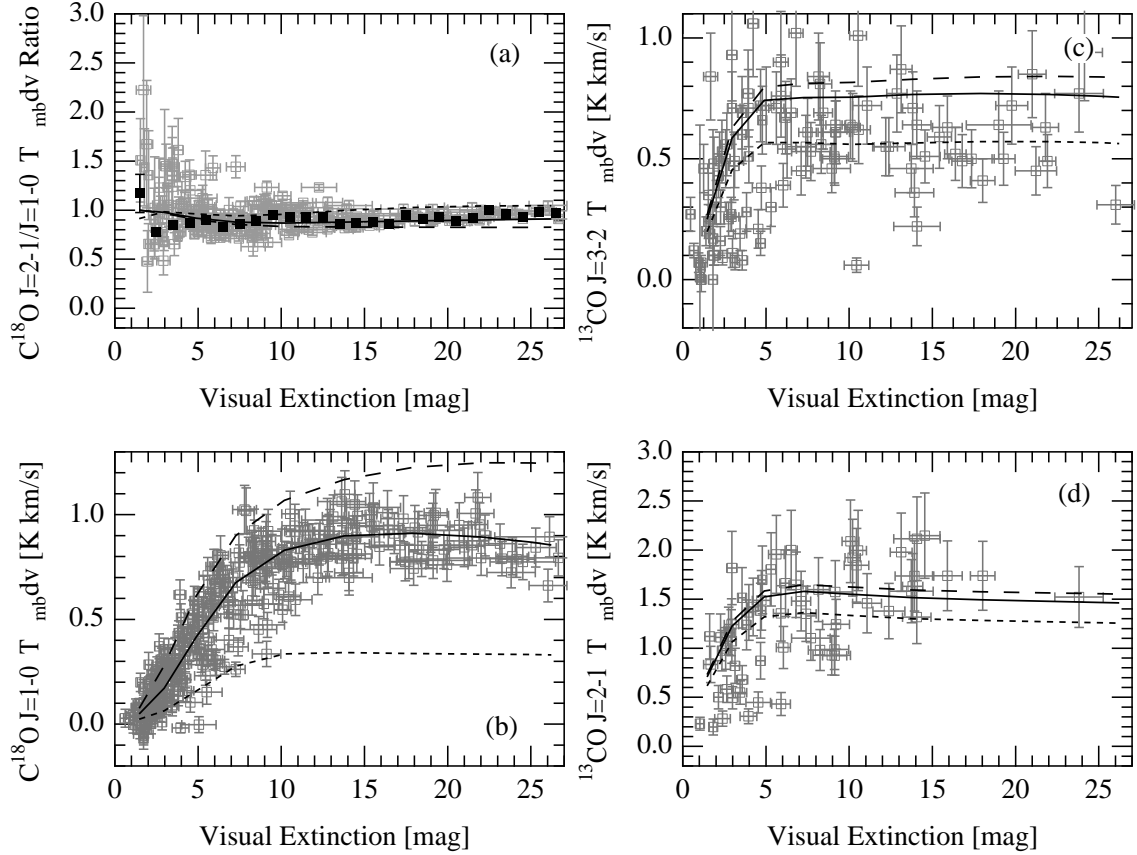}
\caption{
Point by point comparison of 
(a) J$=2-1$/J$=1-0$ line ratio, 
(b) {\rm C$^{18}$O} J$=1-0$ integrated 
intensity, 
(c) {\rm $^{13}$CO} J$=2-1$ integrated intensity 
and (d) {\rm $^{13}$CO} J$=3-2$ integrated intensity 
with A$_V$ for the entire B68 dark cloud.
In these plots the data are presented as open
squares with error bars while solid curves represent the emission
predicted
by a model combining chemistry with a Monte-Carlo radiative transfer
code.  Time dependent model predictions are provided with 
t = $4 \times 10^4$ yr (long dashed line), $1 \times 10^5$ yr (solid line),
and $3 \times 10^5$ yr (short dashed line).   In (a) we include 1 mag binned
weighted average of the J$=2-1$/J$=1-0$ line ratio
}
\label{fig:time_data}
\end{figure}

\begin{figure}
\plotone{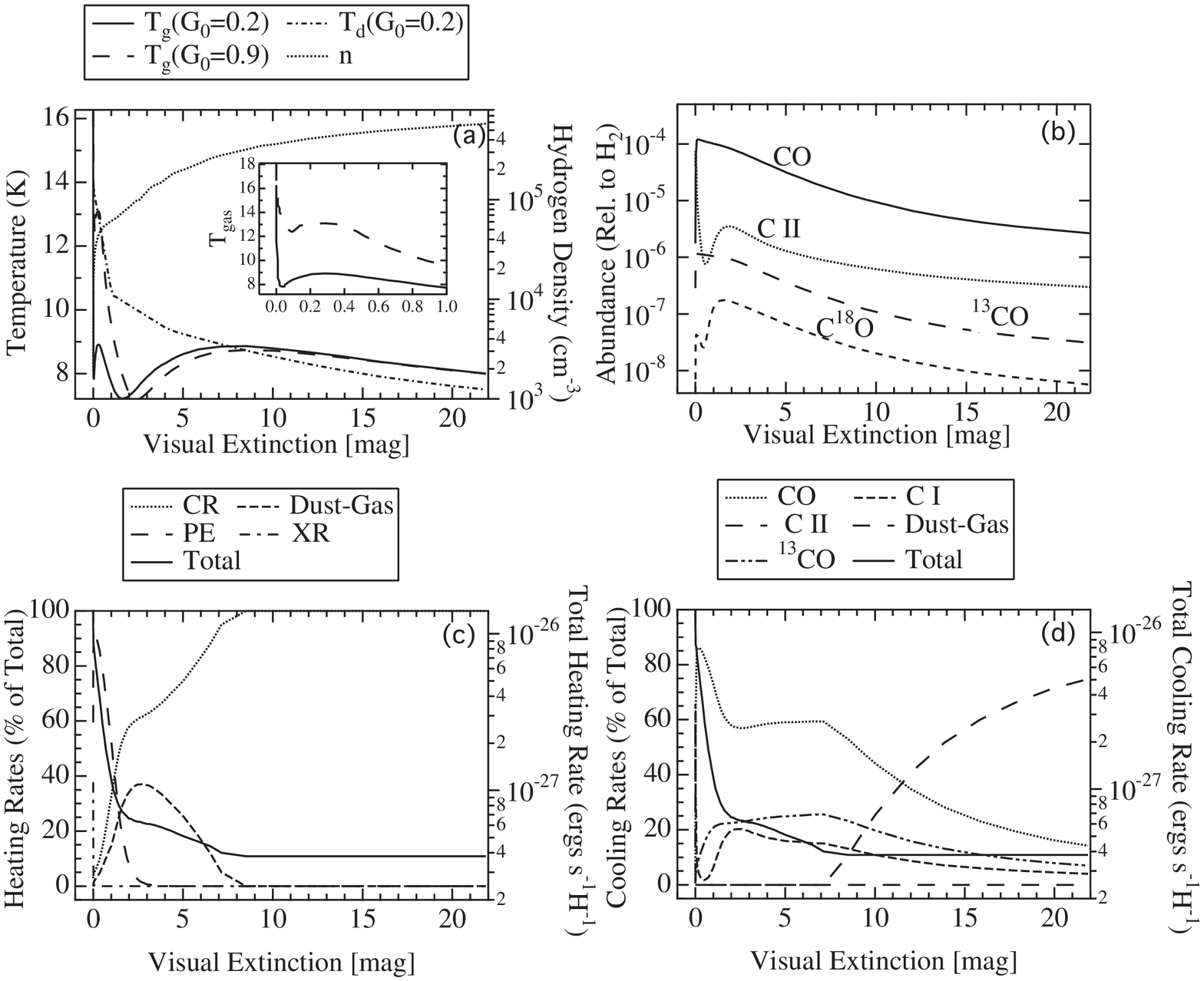}
\caption{
Results from a coupled chemical/thermal model of the B68 dark cloud.
(a) Gas temperature for two models assuming an external
illumination of G$_0$ = 0.2 and 0.9.  Dust temperature structure
taken from a model of B68 of Zucconi et al. (2001).
The gas temperature for the  G$_0$ = 0.2 and 0.9 models
are identical for A$_V > 2^m$.
We do not use a log scale on these figures to emphasize the gas
temperature in layers with $A_V > 0.1$ mag, where the molecular
lines can be used as probes.
(b) Chemical abundances of primary coolants for 
the G$_0$ = 0.2 model.  
(c) Primary cooling contributors as percentage of total.
Total cooling rate is given as a solid line with axis labeled to right.
(d) Primary heating contributors as percentage of total.
Total heating rate is given as a solid line with axis labeled to right.
}
\label{fig:fuv_model}
\end{figure}

In Figure~\ref{fig:fuv_model} we provide a 4 panel plot with
the relevant physical quantities that are inputs (e.g. density, dust
temperature) or outputs (gas temperature, abundance) from the model. 
To aid the discussion,
we also provide the major heating and cooling terms, as a function of 
cloud depth. 
In Fig.~\ref{fig:time_data} we provide the results from 
a series of time-dependent models 
superposed on the observational data for B68.
This series of four panel plots of the \CeiO\ J=2-1/J=1-0 integrated emission
ratio plotted as a function of \av\ (Fig.~\ref{fig:time_data}a), 
along with the integrated
intensity of \CeiO\ J=1-0 (Fig.~1b), 
J=3-2 (Fig.~\ref{fig:time_data}c), and 
\thCO\ J=2-1 (Fig.~\ref{fig:time_data}d) 
as a function of \av\ are used to test our models.  In each panel the
symbols are data points and lines are the model results.

\begin{figure}
\plotone{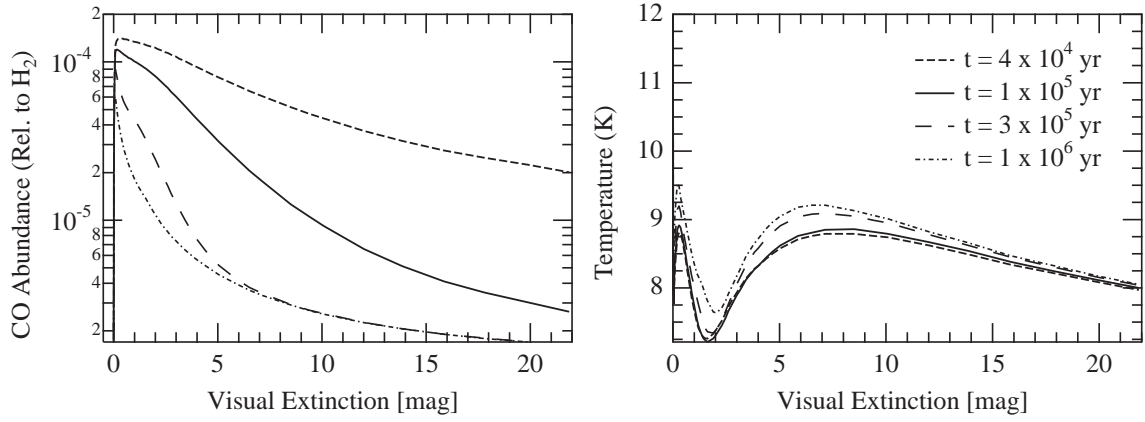}
\caption{
{\em Left:} CO abundance as a function of visual extinction and time.
{\em Right:} Gas temperature as a function of visual extinction and time.
}
\label{fig:time_temp}
\end{figure}

The models shown in Fig.~\ref{fig:time_data} sample a range of times 
and assume a cosmic ray ionization rate of $\zeta =
3.0 \times 10^{-17}$ s$^{-1}$, $\delta_d = 1$, G$_0$ = 0.2.
Our ``best fit'' model
is at a time of t $\sim 10^{5}$ yr, with earlier and
later times showing significant discrepancies with observed emission.   
``Best-fit'' implies agreement with CO
isotopologue data.  As discussed later some modifications to this model are
needed to match additional observations.
The primary effect of time in the models is increasing CO freeze-out,
which lowers the column density as the core evolves.
Thus, the drop in \CeiO\ and \thCO\ column density
directly relates to the reduction in predicted line intensity.  
The magnitude of the decline depends on
the optical depth of the transition, hence it is largest for thinner \CeiO\
J=1--0 emission and smaller for thicker \thCO\ emission lines.   
The \CeiO\ line ratio does not show  strong time-dependence
because CO freeze-out, and the subsequent decrease in CO cooling,
does not produce a sharp temperature rise.  This is demonstrated
in Fig.~\ref{fig:time_temp} where we present the temperature and CO
abundance structure as a function of time.   
As time evolves in regions
with lower density gas with weaker gas-dust coupling (\av\ $<$ 10 mag), 
the depletion of
coolants warms the gas, but the cooling power of
dust-gas collisions compensates for this loss in the densest regions 
(\av\ $>$ 10 mag), a
point first noted by \citet{goldsmith_tg}.
 The CO abundance exhibits a sharp drop
below \av\ $= 1$   mag due to photodissociation and decline at high \av\  due to 
freeze-out.  The CO photodissociation front is confined to a small range of visual
extinction because of the high densities and low UV fields at the boundary layer.

Overall,
the gas temperature
shows structure but remains within a narrow 
range of 7 -- 9 K.
The structure is due to the
interplay between the rise and fall of varied heating and
cooling agents.  
At very low \av\ (below $\sim$ 0.1 mag) the sharp drop of gas
temperature is due to the rise of CO as a gas-phase coolant.
The gas
temperature shows a slight increase at \av\ $\sim 0.1$ mag as the cooling
power of C$^0$ and C$^+$ is lowered as a direct result of CO
formation.   Between \av\ 0.2 to 1.7 mag the temperature declines due
to the decreasing efficiency of photoelectric heating.  The decline is
reversed for \av\ $> 1.7$ mag when the sharp density increase,
higher dust temperature, and CO freeze-out combine to
produce a rise in dust-gas collisional heating with a reduction in
cooling power that compensate
for the loss of photoelectric heating.  We note that the gas temperature 
is low between \av\ = 1 -- 4 mag precisely because CO is present as a coolant.
When T$_d$ $<$ T$_g$ (\av\
$\sim 7.5$ mag) dust collisions become gas coolants
producing a slow temperature decline towards the core
center.  In these models the gas and dust are never completely coupled.

\subsubsection{Turbulent Velocity Width}

\begin{figure}
\plotone{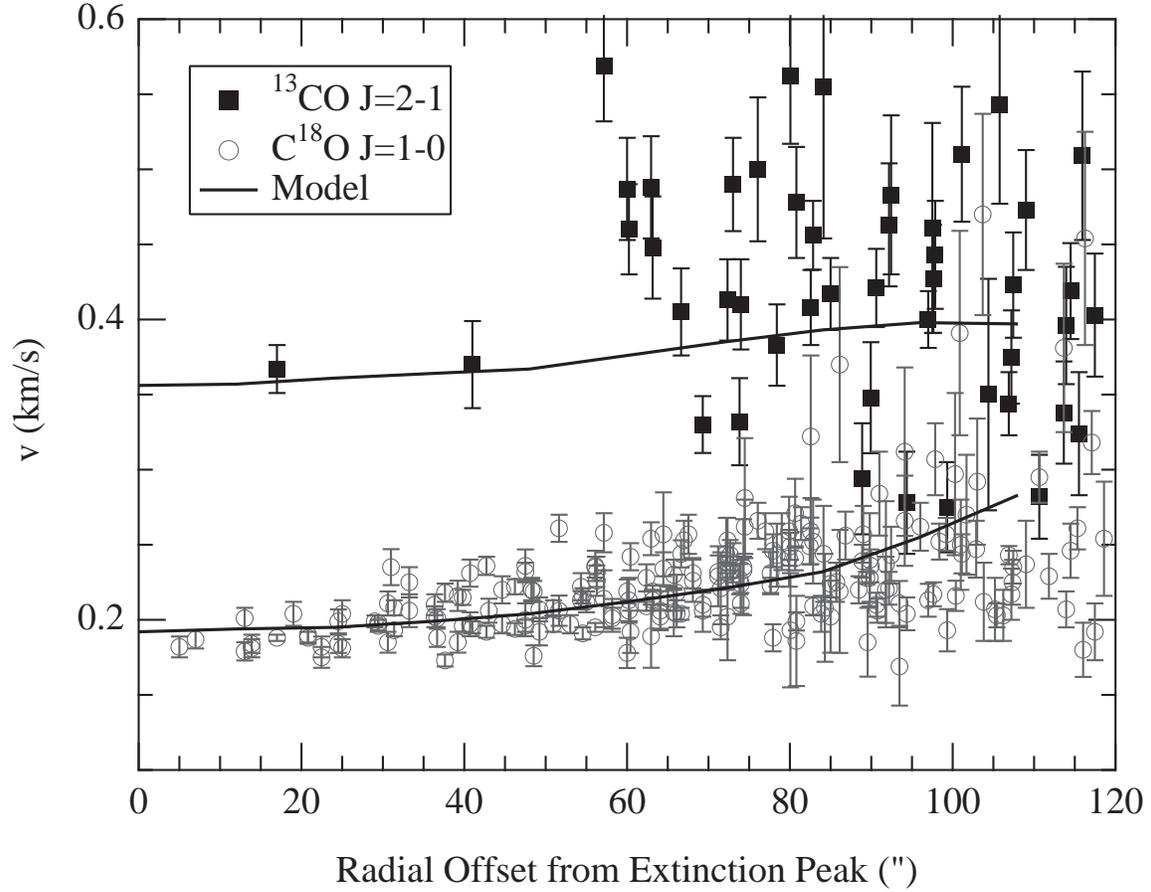}
\caption{
Velocity linewidth (FWHM) determined by fitting \CeiO\ J=1-0 and \thCO\
J=2-1 observed and model spectra with Gaussians shown as a function of 
radial distance from the extinction peak.  
Data are given as points and model results from the best fit solution 
as a solid line.  
}
\label{fig:dv}
\end{figure}

Fig.~\ref{fig:dv} presents the spectral line full
width at half-maximum (FWHM) as a function of radial distance from the
extinction peak (i.e. the assumed center of gravity).   For the data in this
plot the velocity FWHM is determined by Gaussian fits 
to each spectrum with a $\ge 5\sigma$
emission line detection.  The  variation of \CeiO\ J=1-0 
linewidth with position has
been discussed previously by \citet{lada_b68}.  However, the \thCO\
J=2-1 profiles are much wider and exhibit more scatter.   

Our best fit model (solid lines) faithfully reproduces the variation in
both \CeiO\ and \thCO .
The linewidth difference between these two species can be interpreted 
by the \thCO\ J=2-1
emission becoming optically thick in outer layers with higher velocity
dispersion.  We estimate $\tau \sim 5$ for \thCO\ J=2--1 (from the
observed \CeiO /\thCO\ J=2--1 line ratio and $^{12}$C/$^{13}$=65) and
$\tau \sim 0.7$ for \CeiO\ J=2--1.
With lower opacity \CeiO\ emission traces a larger volume dominated by
layers with lower turbulence.
Thus in comparative chemical studies the variation in the velocity
linewidth can contain information on the depth that is
being probed.

\subsection{Constraints on the External FUV Radiation Field}
\label{sec:uvfield}

\begin{figure}
\plotone{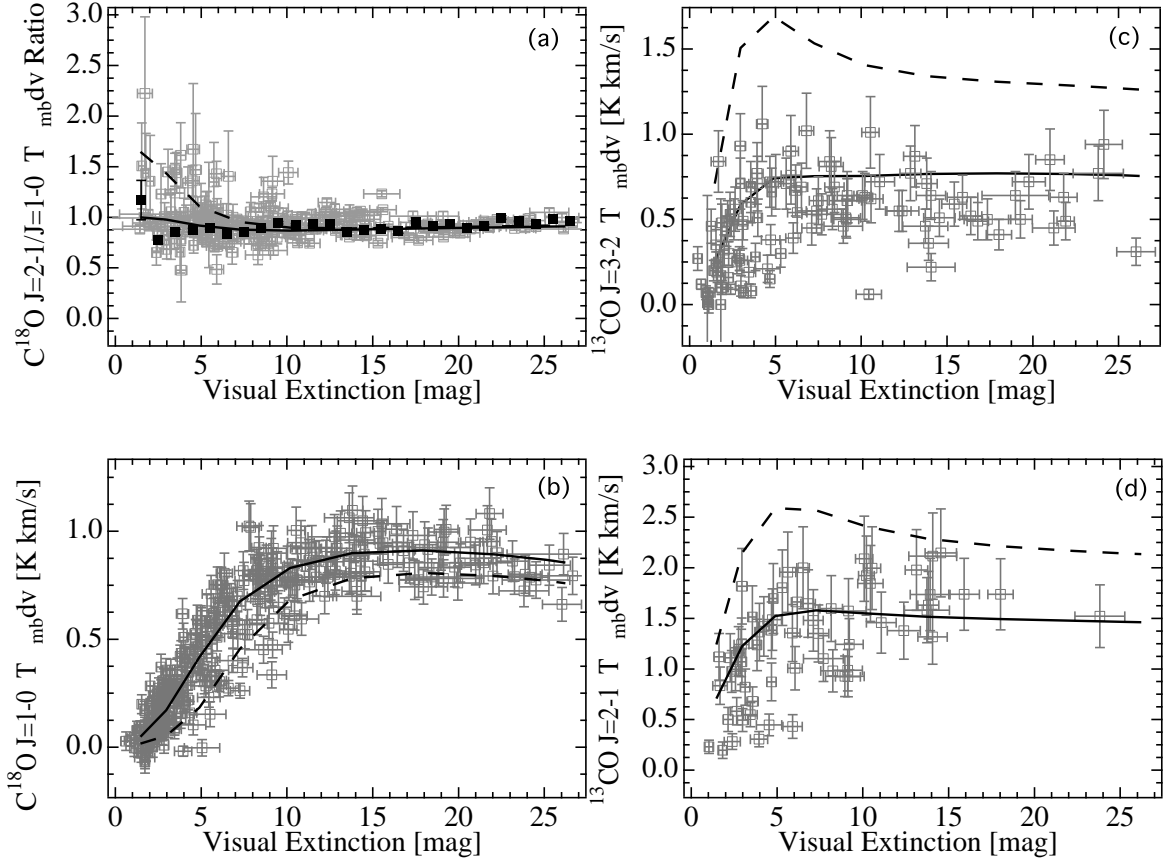}
\caption{
Comparison of chemical/thermal models with different values of the
radiation field.
(a-d) as in Fig.~\ref{fig:time_data}.
Two values of the external radiation field are examined:
\go\ = 1 (dashed lines) and  \go\ = 0.2 (solid lines).
}
\label{fig:uv_data}
\end{figure}

As noted earlier the model can not match these data assuming a core bathed
within the standard interstellar UV radiation field.   This is
demonstrated in Fig.\ref{fig:uv_data} where we present models with
two different values of \go .  We note that other values of \go\ were
examined and present here two extremes;  only values of \go\ 
$\lesssim 0.2$ can match our data.
The \go\ = 0.2 model, which is our
best fit solution, matches all data to within the errors.  
The \go\ = 0.9 model fails on two accounts. (1) Enhanced \CeiO\
photodissociation lowers the abundance,
resulting in an inability, {\em at any time}, to reproduce the \CeiO\
integrated emission. 
(2) The gas temperature (see Fig.~2a) in the outer layers is warmer,
resulting in stronger \thCO\ emission than observed.
An additional demand 
for a colder cloud is  \CeiO\ J=3-2 emission, which 
was not detected on two separate
occasions to a level of $T_{MB} \sim$1.2 K (3$\sigma$; \S~\ref{sec:obs}).  
Our best-fit model produces an peak intensity for this line of 1.6 K, which
along with \thCO\ J=3--2,
may indicate that this model is still too warm.  
In sum, \thCO\ data, and \CeiO\ line ratios,
suggest temperatures that are quite cold, $\lesssim 8$ K, 
even in the layers with relatively low extinction (\av\ = 1 mag).  
The temperature is below the assumed dust temperature at these depths.  

One issue with our model and the predicted strong \thCO\ emission
for \go\ = 0.9 is that the velocity field is treated as static.
Contributions from the line of sight velocity field to the line
profile have been accounted for within the increasing turbulent contribution
to the velocity width (\S~\ref{sec:model}).  B68 has clear
evidence for rotation which induces line shifts of up to 50\% of the
linewidth across the face of the cloud.  If rotation were included
in the model it could potentially lower the opacity of the optically
thick \thCO\ line emission.  We have investigated this possibility using the 2-D
Monte-Carlo radiation transfer code of \citet{hgvdt} and the rotation rate of 4.77 km
s$^{-1}$ pc$^{-1}$ estimated by \citet{lada_b68}.  The rotation axis of B68 shows a small
6 degree slant which may be indicative of some tilt along the line of sight.
However, since this is small we have assumed the rotation is oriented along the plane of
the sky.  We find that the effect of rotation on the \thCO\ emission is small with only a
10\% change in the opacity.  Thus, rotation cannot be an answer for the weak \thCO\
emission.

In addition, rotation cannot account for
optically thin \CeiO\ J=1-0 emission, where the integrated intensity depends strongly on
its abundance.  Rather, the photodissociation rate
(with self-shielding) plays a major role in determining whether enough
\CeiO\ exists to emit at appreciable levels.  
Moreover, the non-detection of optically thin
\CeiO\ J=3-2 emission suggests that the layers traced by these species
need to be cold with $T_{gas} <$  8 K.

Another tracer of the external FUV field is 
optically thick $^{12}$CO emission.  In B68 the J=1-0 transition has been
detected with a line peak intensity of $T_{mb} =$ 6.7 K 
(F. Bensch, priv. comm.; KOSMA).  Our \go\ = 0.9 model predicts 
$T_{mb} \sim$8.7 K, while our best fit model predicts $T_{mb} \sim$2.7 K.   
We note that the external temperature in our best fit
model is 16 K, but that this warm gas does not fill significant volume.
Given this
trend our models clearly rule out an external field as strong as the
\citet{draine78} UV field (\go\ = 1.7) and argue for something smaller
than the \citet{habing68} field.  This is in agreement with the study by
\citet{young_tgas} who also found low UV fields are needed to match CO
emission in three pre-stellar cores  embedded in the Taurus Molecular Cloud, although they also found evidence for high external
temperatures of 10 -- 14 K.

\subsection{Model Sensitivity and Inner Core Gas Temperature}

\subsubsection{Cosmic Ray Heating}

\begin{figure}
\plotone{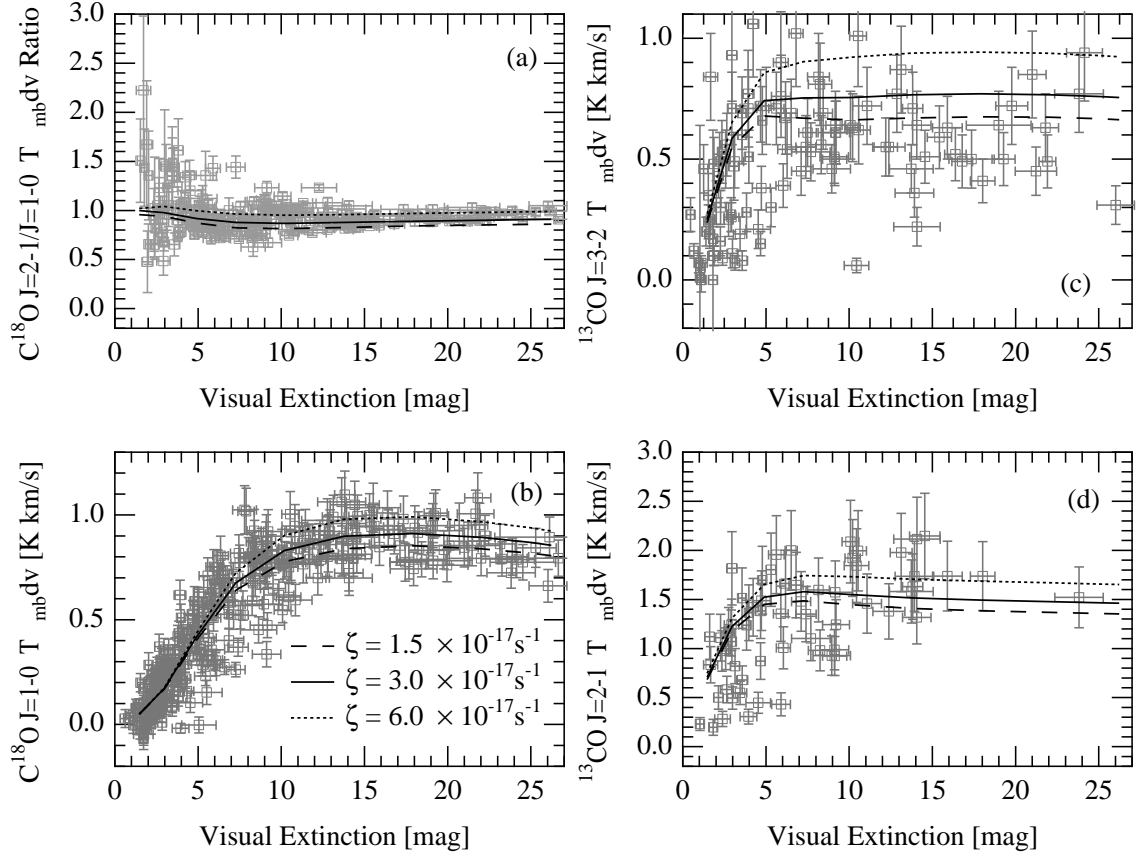}
\caption{
Comparison of chemical/thermal models with different values of the
cosmic ray ionization rate with labels provided above.  
(a-d) as in Fig.~\ref{fig:time_data}.
}
\label{fig:cr_data}
\end{figure}

\begin{figure}
\plotone{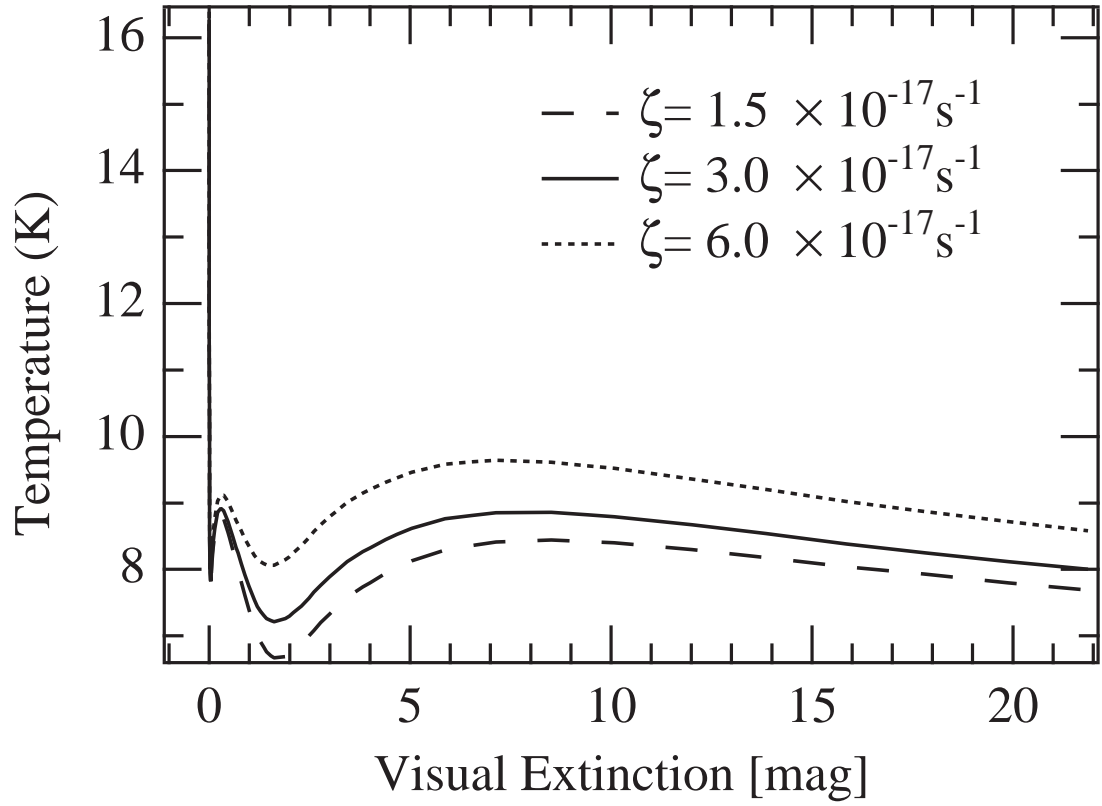}
\caption{
Derived temperature structure for models with
different values of the
cosmic ray ionization rate with labels provided above.  
}
\label{fig:cr_temp}
\end{figure}

The chemical/thermal model does exhibit moderate sensitivity
to the assumed level of cosmic
ray heating.  In Fig.~\ref{fig:cr_data} we provide a comparison of the
observations to models with three different values of the 
cosmic ray ionization rate.  Fig.~\ref{fig:cr_temp} provides the derived
temperature structure from each model. Some temperature differences are found
for \av\ $>$ 1 mag, the depth where cosmic ray heating becomes
significant when compared to photoelectric heating
(Fig.~\ref{fig:fuv_model}).
Models with $\zeta \sim 1.5-3.0
\times 10^{-17}$ s$^{-1}$ provide reasonable matches to the data.  In
contrast, when $\zeta > 6.0 \times 10^{-17}$ s$^{-1}$ the gas temperature in
depths traced by \CeiO\ and \thCO\ rises to a point where the
model/observation comparison is not as favorable.  Given the
assumed observational errors, our models find reasonable fits with
$\zeta = 1 - 6 \times 10^{-17}$ s$^{-1}$.   The derived cosmic-ray
ionization rate is consistent with the value estimated by
\citet{vdt_cosmicray} and \citet{webber_cosmicray}. 

\subsubsection{Dust-Gas Coupling}
\label{sec:dg}

Since CO and its isotopologues are heavily depleted in abundance in the
central core it is worth examining whether our model sets any
constraints on the central core temperature.  This is important because
observations of \NHthree\ in B68 suggests gas temperatures $\sim$10--11 K
\citep{hotzel_nh3, lai_nh3}.   Ammonia likely forms from 
\NtwoHp\ \citep{aikawa_be}, which itself only forms in abundance when CO freezes onto
grains.  Thus ammonia is chemically selected to probe the gas in regions where CO is
losing sensitivity.
This suggests that the temperature deeper in the cloud may be warmer
than the $\sim 8-9$ K in our best-fit model.  

\begin{figure}
\plotone{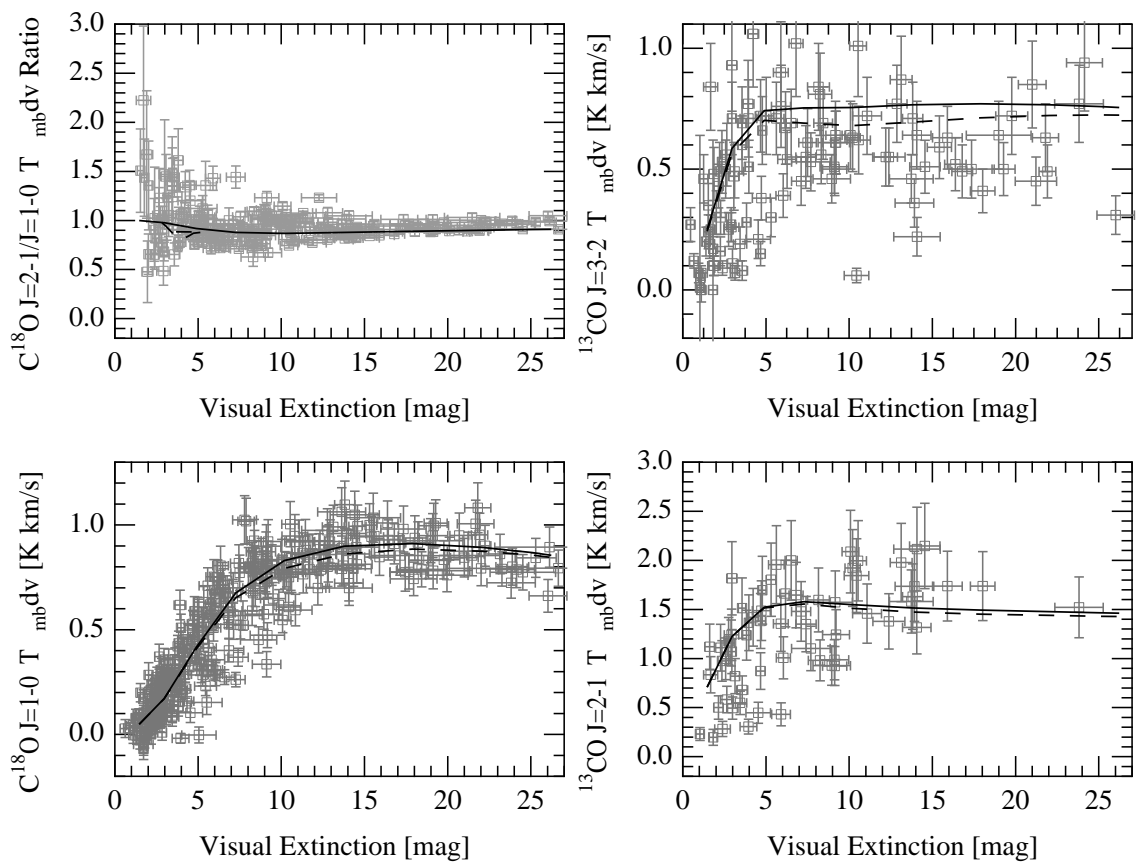}
\caption{
Comparison of chemical/thermal models with variable degrees of dust-gas thermal
coupling. 
(a-d) as in Fig.~\ref{fig:time_data}.
Two values of $\delta_d$ are examined: 
$\delta_d = 1$ (solid line) and $\delta_d = 0.1$ (dashed line).
}
\label{fig:dg_data}
\end{figure}

\begin{figure}
\plotone{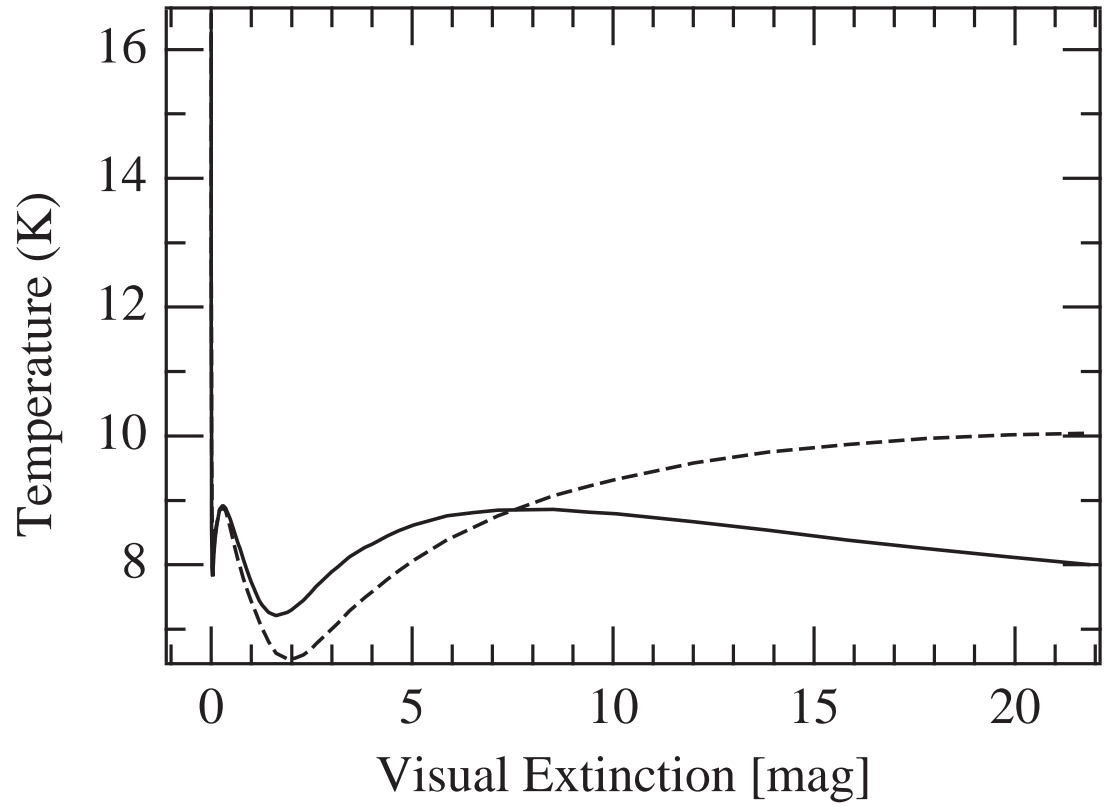}
\caption{
Derived temperature structure for models with
variable degrees of dust-gas thermal
coupling.
Two values of $\delta_d$ are examined: 
$\delta_d = 1$ (solid line) and $\delta_d = 0.1$ (dashed line).
}
\label{fig:dg_temp}
\end{figure}

Fig.~\ref{fig:dg_data} and Fig.~\ref{fig:dg_temp} provide the comparison
of theory to observations and the temperature profiles for models where
we vary $\delta_d$ (the ratio of dust abundance in modeled region to
that of the diffuse ISM).   We find that to reach T$_{gas} \sim 10$ K,
in the central core, as inferred from NH$_3$ emission line ratios,
requires an order of magnitude reduction in the
dust-gas thermal coupling rate ($\delta_d = 0.1$).   This has a small 
effect on the \CeiO\ and \thCO\ emission demonstrating that
their emission is not a probe of  the  inner core gas temperature.
Curiously lowering the dust-gas thermal coupling rate 
reduces \CeiO\ and \thCO\ emission.  This is because these two species
are predominantly tracing depths where collisions with dust are
heating the gas (see Fig.~\ref{fig:fuv_model}). 
 Thus, provided  our understanding of dust-gas thermal coupling rates is correct, to match
NH$_3$ derived gas temperatures requires a reduction in the rate, potentially due to
changing dust properties.

The conclusion regarding weaker gas-dust coupling rests upon proper
modeling of the dust temperature and the calibration of the ammonia 
thermometer.  In our models we
have not directly examined dust emission and 
used the formula provided by \citet{zucconi_td} with some
modifications to account for reduced radiation field (\S 3).  
Our modifications to the dust temperature did result in reducing the dust
temperature at high depth into the cloud. 
However, at \av\ $>$ 10 mag our ``best-fit" solution assumes 
$T_{dust} \sim$8 K, which
is quite close the value for similar depths used by
\citet{galli_tg} in their core stability study of B68 and is also
similar to the value inferred by \citet{zucconi_td} and \citet{evans_td} 
for the inner radii of Bonnor-Ebert spheres.  Thus our the requirement
for reduced dust-gas coupling would not be altered with more exact
treatment of the dust emission. 

One question is whether the ammonia thermometer and measurements are
calibrated to 
an accuracy of $\sim 2$ K.  
The \citet{hotzel_nh3} measurement of the gas temperature from ammonia
emission taken with a similar resolution (40$''$) to our data.  The
raw error of this measurement is 9.7 $\pm 0.3$ K, which is increased to
1.0 K allowing for 20\% differences in the filling factor and T$_{ex}$
between the two lines.  This seems overly generous given the similarity
in critical density and filling factor between these lines.  Moreover
the \citet{lai_nh3} temperature estimate is 10.8 $\pm 0.8$ K.  Thus both
analyses suggest at
least a 3$\sigma$ difference between the central core gas temperature in
our model of $\sim 8$ K and measurements.
Regarding the calibration of the ammonia thermometer, 
for T $> 10$ K the observed
(2,2) and (1,1) inversion transition rotational temperature is not equivalent 
to the gas kinetic temperature.   However, close to 10 K the 
rotational temperature is a measurement of T$_k$ without any 
correction \citep{danby_nh3}.
Thus, provided this calibration is accurate,  the observed difference is 
suggestive that the dust-gas coupling needs to be reduced.
 
\begin{figure}
\plotone{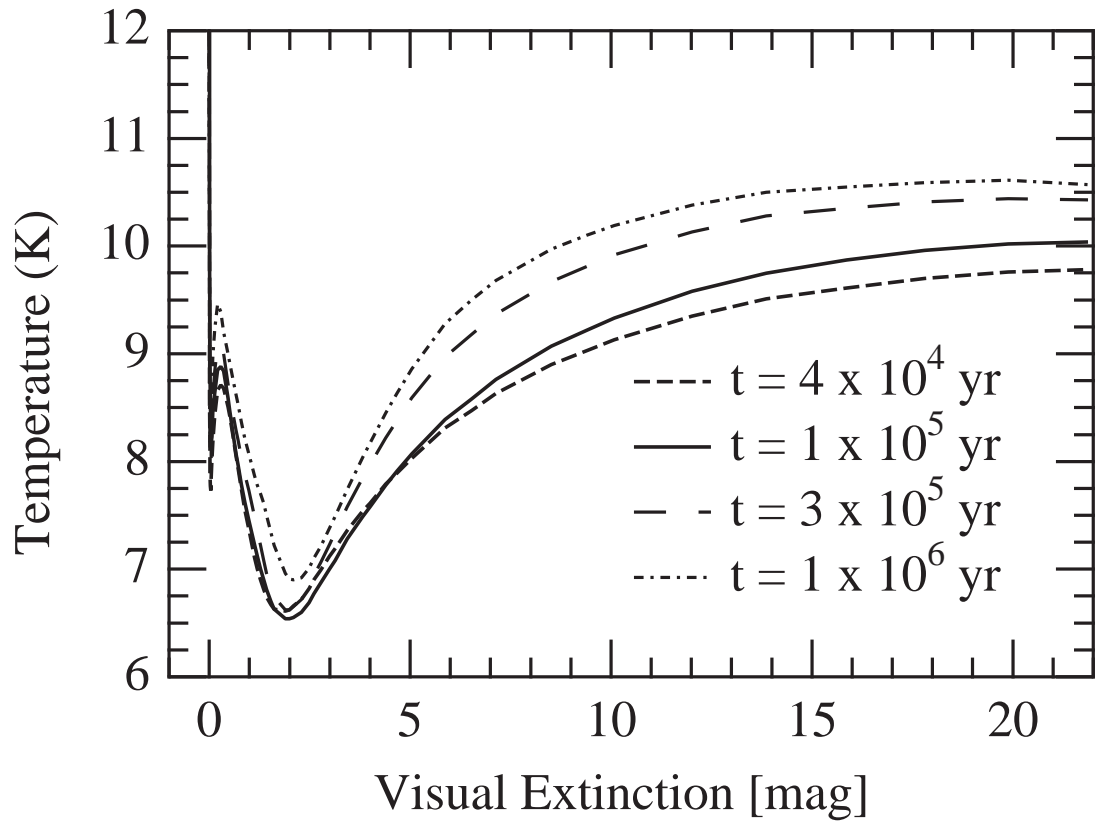}
\caption{
Derived temperature structure as a function of time for models assuming
$\delta_d = 0.1$. 
}
\label{fig:dg_time}
\end{figure}

Given the need for lower dust-gas thermal coupling
in Fig.~\ref{fig:dg_time} we re-examine the time evolution of the gas
temperature in a model with reduced coupling.  
Similar to the discussion in \S~\ref{sec:compobs}, we find that as time
evolves and gas coolants deplete the gas temperature rises.  In this
case the effects of freeze-out on $T_{gas}$ are stronger, but at most
depletion warms the core by $\sim 1$~K.  The temperature is actually
lower between 0 $<$ \av\ $< 7$ mag, because the main heating mechanism
is dust collisions in this region.   Thus, while we have found that
freeze-out does raise the temperature, these effects are small and we
remain in agreement with previous studies of the gas temperature in cores
dominated by freeze-out \citep{goldsmith_tg, galli_tg}.

\section{Discussion}

\subsection{The Meaning of Time in these Models}

In our models we find a ``best-fit'' to the observations at 
t $\sim 10^5$ yr.  This does
not suggest that we constrain the age of this object to be $10^5$
yrs.   We have used a density profile that is fixed in time in our
time-dependent chemical calculations.  Thus we have not accounted for contemporaneous 
density evolution along with the chemistry.  To compensate for this
effect we have assumed that all CO is pre-existing in the gas at t = 0
in our chemical network solutions.  With these assumptions this time is
not the time since the cloud became molecular (since CO is pre-existing)
nor is the time since the density structure evolved (since the structure
is fixed).  Rather, the best-fit time represents the
time where sufficient pre-existing CO has been frozen onto 
grains and the predicted
gas emission matches observations.  
We have also assumed a sticking coefficient of unity. A lower
value would result in a higher age determination.
 Moreover, if we had included any the effect of
dust coagulation on the chemistry there would be an increase
in the derived timescale.
In this sense our derived time is
only a {\em lower limit} to the true age of B68.

We have included a mechanism for
returning CO to the gas via constant cosmic ray 
desorption of CO ice, which does
depend on our assumed binding energy.  In our models we do not reach a
time where desorption is relevant in the sense that at late times when
pre-existing CO is frozen and equilibrium is reached between desorption
and depletion there is not enough CO in the gas to reproduce the
observed emission.
The binding energy would need to be lowered well below measured values
\citep{collings_cobind}
and the rate of CO desorption increased above estimated levels
\citep{bringa} to change this condition and raise our time estimate.

\subsection{Grain Evolution}
\label{sec:grainevol}

Our multi-molecular comparative chemical study (e.g. CO and \NHthree ) suggests
that the dust-gas coupling rate is reduced in the center of B68. 
This  potentially represents evidence for grain coagulation
inferred from gas observations.  In the literature there are several lines of
evidence for grain evolution from dust observations.  
These include variations
in the ratio of visual extinction to reddening, $R_V$,
\citep[][and references therein]{fitzpatrick04} which have
been attributed to changes in the grain size distribution
\citep{kim_martin94, weingartner_draine01}, studies of the far-infrared
emission which find a lack of small grains in denser regions
\citep{stepnik_pronaos}, and the requirement for enhanced sub-mm
emissivity in star-forming regions \citep{vdt_gl2591, evans_td}. 
Theory of grain coagulation suggests significant grain evolution 
is expected in only a few million years \citep{weidenschilling_ggrowth}, the 
expected lifetime of the pre-stellar phase \citep{lee_myers_survey}.

To examine what grain evolution (e.g. coagulation) is 
needed to reach $\delta_d = 0.1$ we have used the 
MRN \citep{mrn} grain size distribution.  This is given by:

\[
dn_{gr} = Cn_Ha^{-3.5}da, \;\;\;a_{min} < a < a_{max},
\]

\noindent with $a_{min} = 50$ \AA\ and $a_{max} = 2500$ \AA . \citet{draine_lee}
find the normalization for silicate grains to be $C = 10^{-25.11}$
cm$^{2.5}$.  Integration of this equation provides $n_{gr} = 1.75 \times
10^{-10}n_H$ for standard ISM grains.   To obtain $\delta_d = 0.1$ (while
maintaining a$^{-3.5}$)
necessitates a minimal change in
the size of the smallest grains, to $a_{min} = 150$ \AA .   This analysis
is simplistic as we have not discussed extensions of MRN towards even
smaller grains that are needed for the ISM in general \citep{desert_pah,
weingartner_draine01}.  However, it is illustrative that only small
changes are needed in the lower end of the size distribution to account
for the reduction in grain surface area demanded by available 
gas emission data.  


One inconsistency is that we have included small grain photoelectric heating  
in our models and need to have a warmer central core with
less small grains to match \NHthree\ observations.   To this end we
investigated the necessity for  photo-electric heating in our
theory/observation comparison.
We find that a reduction of the photoelectric heating rate
of more than 50\% does not agree with our data (even with a compensating 
increase in cosmic ray heating).\footnote{In principle the dust-gas
coupling could be stronger at the low density core edges because of
warmer dust temperatures and the presence of small grains and weaker in
the center due to the lack of small grains.  
This might further compensate from the loss of very small
grains and allow for a greater decrease in photoelectric
heating.} Thus, based upon the adopted formalism 
for photoelectric heating and dust-gas thermal coupling, we 
require some small grains to be
present in the outer layers but these same grains are absent in the
inner core.  This is not unreasonable given the large
density increase seen in B68
\citep[Fig.~\ref{fig:fuv_model};][]{alves_b68}.

 It is worth noting that \citet{bianchi_td}  find no evidence for dust
coagulation based on measurements of the sub-millimeter emissivity and
comparison to available grain models.  However, the measured value is higher than
that of diffuse dust, although only at the 2$\sigma$ level.   
 
\subsection{Comparison to Previous Results}
\label{sec:modcomp}

A major issue with our suggestion that  the UV field is reduced is 
dust emission models which require a core exposed
to ISRF derived by \citet{mmp}, which has \go\ = 1.6 \citep{draine_li01}.  
Moreover the intensity of the ISRF incident on B68 has been estimated  
independently by \citet{galli_tg}  using a variety of dust continuum emission observations with
no evidence for any reduction in field strength.
Thus there is nearly an order of magnitude difference in 
UV field strength between gas and dust models.

 A likely possibility to reconcile these differences is that there exists an extended layer of
diffuse gas around B68 that might unassociated with the core.
For instance, Lombardi et al. (2006, in prep.) have used the Near-IR extinction 
method to map the extinction structure within the Pipe nebula, 
in which B68 is embedded.  This map shows a large degree of extended extinction in this 
cloud,
which provides additional UV shielding for B68.  If this layer consists  of only
1 mag of extinction, then the strength of the local radiation field for B68 would be 
increased by a factor of 1/exp(-3) or $G_0$ = 2.  Moreover,  any H$_2$ in this layer can
contribute to CO self-shielding \citep{bensch_pdr}, requiring even less dust extinction.  
This would allow for higher UV fields in
agreement with estimates from the dust emission.  
 We investigated the inclusion of such a halo in
our models and find little change to our results for CO isotopologue emission. 


\subsection{Cloud Support}

\citet{lada_b68}
used the observed linewidths of \CeiO\ and \CtfS\ to derive an estimate
of the thermal to non-thermal (turbulent) pressure of the B68 core,
which is given as:

\[
R = \frac{a^2}{\sigma_{nt}^2}
\]

\noindent In this expression $a$ is the sound speed of the gas and $\sigma_{nt}$ is
the three dimensional velocity dispersion.  This measurement used
fits to the observed emission profiles that are an integration along 
the line of sight to
demonstrate that B68 is dominated by thermal pressure.  
In our analysis we have derived a fit to the velocity structure in B68
(Fig.~\ref{fig:dv}) which, for a static model, includes contributions
from the thermal and non-thermal velocity dispersions as a function of
position in the core.\footnote{We note
that this analysis includes no contribution from any systematic cloud
motions and thus the derived non-thermal contribution is an upper limit.}

\begin{figure}
\plotone{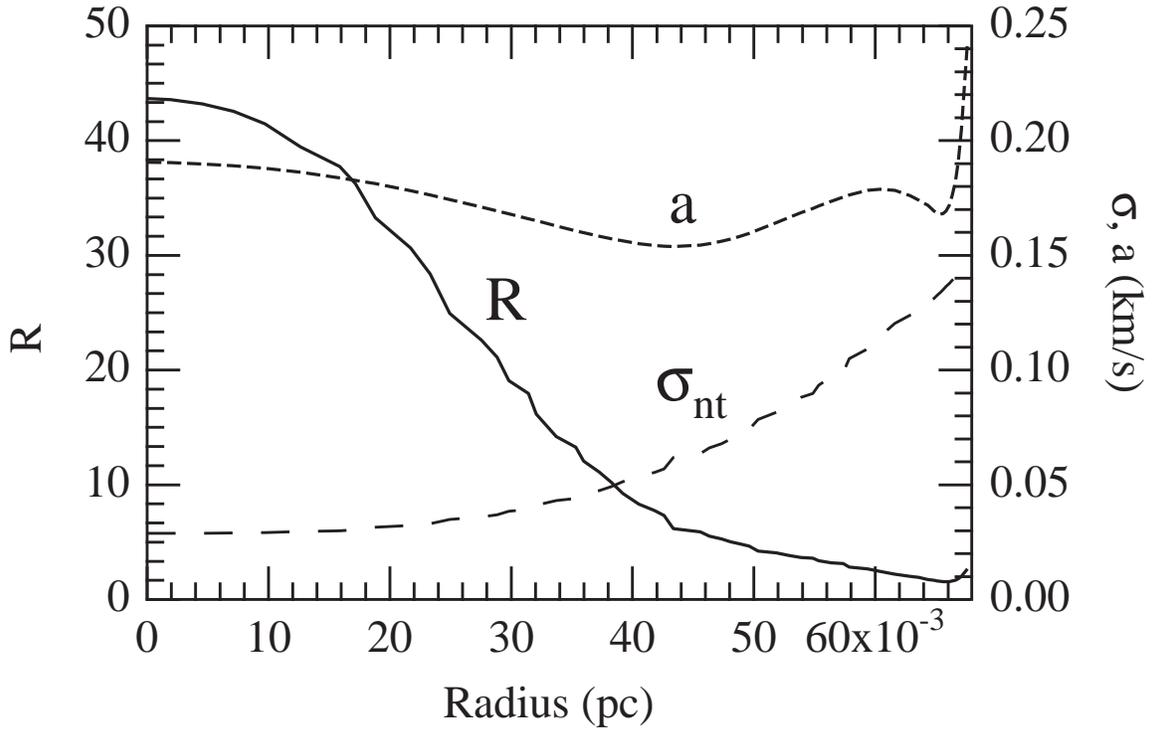}
\caption{
Examination of the derived velocity and pressure structure in B68.
Left axis shows the value of R, the ratio of thermal to non-thermal
pressure, as function of radius.  The right axis refers to the values of
the thermal (a) and non-thermal ($\sigma_{nt}$) velocity dispersion
also given as a function
of radius.
}
\label{fig:velstruct}
\end{figure}

In Fig.~\ref{fig:velstruct} we show how the the
thermal to non-thermal pressure ratio varies as a function of radius 
in B68.  There is a sharp variation throughout the object with the center
completely dominated by thermal motions and increasing turbulent
contributions towards larger radii.
At all positions the core is dominated by thermal pressure, confirming
the previous result.
\citet{lada_b68} derived values of $R \sim 4-5$ for the inner core and
$R \sim 1-2$ for the edges.  In our model the peak \CeiO\ abundance is
found at r $\sim 0.04$ pc, a point which corresponds to $R \sim 5$,
close to the value inferred from observation.
Thus, in cores with significant abundance  and velocity field structure,
the line of sight average velocity dispersion from the emission profile 
is heavily weighted to radii which have the largest abundance.  

It is worth noting that B68 is dominated by thermal pressure, 
but it is not an isothermal object.  In our preferred model
that matches both CO isotopologues and \NHthree\ ($\delta_d  = 0.1$;
\S~\ref{sec:dg}), $>$ 95\% of the enclosed mass lies in layers with a
gas temperature variation of $\sim 3$ K.  These changes are within the
temperature range examined in  the thermal balance and stability 
study of B68 by \citet{galli_tg}.  
However, the sense of the gradient is opposite to what is normally
examined for a pre-stellar object  (i.e. the temperature increases
inwards). Nonetheless these changes are small and it is likely that the
conclusions of \citet{galli_tg} are still relevant and these objects
will have similar characteristics to isothermal models.

\section{Conclusions}

We have combined a multi-molecular and multi-transitional observational
study and a thermal/chemical model to examine
the line of sight gas temperature structure of the Barnard 68 core.
Our primary conclusions are as follows.

\noindent (1) The gas temperature is cold $\sim 6-7$ K in the layers
traced by \thCO\ and \CeiO\ and is warmer $\sim 10$ K in the gas 
deeper within the cloud probed by \NHthree .  For much of the
cloud mass the temperature gradient is small $\sim 3$ K, but increases
towards the center.  With significantly more observational constraints
than previous work, this 
represents the best determination of the gas
temperature structure in pre-stellar cores.

\noindent (2) In agreement with previous studies \citep{hotzel_b68,
bergin_b68} we find evidence for
large-scale freeze-out of both CO isotopologues and demonstrate that B68
is dominated by thermal pressure.

\noindent (3) To match the data to a model we require a UV field
that is weaker than the standard IRSF.  This is in conflict with a previous 
examination of the dust emission and we discuss how this discrepancy can be resolved.

\noindent (4) We find that the dust-gas coupling rate 
must be reduced by nearly an order of magnitude, potentially through a
lack of small grains in the densest regions.  This presents an
argument from gas observations for grain coagulation in the central
regions of the core.

 Our detailed comparison of molecular emission to a PDR model in a low UV environment finds some
differences which can be reconciled by assuming that the physics of dust-gas coupling are represented correctly by changing grain properties.  Whether these results have general applicability
requires more detailed investigations of other sources with well characterized density structure,
preferably with high UV fields.
Additional improvements would result from more direct modeling of \NHthree\ emission.

\acknowledgements

We are grateful to the referee for a thorough and detailed report which
improved this paper.
This work is supported by the National Science Foundation 
under Grant No. 0335207.  EAB is grateful to discussions with F. Bensch 
to D. Lis, M. Tafalla, M. Gurwell, and G. Herczeg for various aspects of
data calibration and observing.

\bibliography{ted}

\newpage

\begin{deluxetable}{lc}
\label{tab:initabun}
\tablecolumns{2}
\tablewidth{2in}
\tablecaption{Initial Abundances}
\tablehead{
\colhead{Species} &
\colhead{Abundance\tablenotemark{a}}}
\startdata
He  &    0.14            \\
H$_2$O$_{ice}$& 2.2 $\times 10^{-4}$\\
H$_2^{18}$O$_{ice}$ &4.4 $\times 10^{-7}$ \\
CO &8.5 $\times 10^{-5}$\\
$^{13}$CO &9.5 $\times 10^{-7}$\\
C$^{18}$O& 1.7 $\times 10^{-7}$\\
Fe& 3 $\times 10^{-8}$\\
Fe$^+$ & 1.2 $\times 10^{-11}$\\
\enddata
\tablenotetext{a}{Abundances are relative to total H.}
\end{deluxetable}
  
\newpage


\end{document}